# On-Stack Two-Dimensional Conversion of MoS$_2$ into MoO$_3$

Taeg Yeoung Ko,[1] Areum Jeong,[1] Wontaek Kim,[2] Jin Hwan Lee,[3] Youngchan Kim,[3] Jung Eun Lee,[1] Gyeong Hee Ryu,[4] Zonghoon Lee,[4] Min Hyung Lee,[1] Changgu Lee[3] and Sunmin Ryu[2,5]\*

[1]Department of Applied Chemistry, Kyung Hee University, Yongin, Gyeonggi 446-701, Korea

[2]Department of Chemistry, Pohang University of Science and Technology (POSTECH), Pohang, Gyeongbuk 790-784, Korea

[3]School of Mechanical Engineering, Sungkyunkwan University, Suwon, Gyeonggi 440-746, Korea

[4]School of Materials Science and Engineering, Ulsan National Institute of Science and Technology (UNIST), Ulsan, 689-798, Korea

[5]Division of Advanced Materials Science, Pohang University of Science and Technology (POSTECH), 50 Jigokro-127, Pohang, Gyeongbuk 790-784, Korea

\*E-mail: sunryu@postech.ac.kr

**Abstract**

Chemical transformation of existing two-dimensional (2D) materials can be crucial in further expanding the 2D crystal palette required to realize various functional heterostructures. In this work, we demonstrate a 2D "on-stack" chemical conversion of single-layer crystalline MoS$_2$ into MoO$_3$ with a precise layer control that enables truly 2D MoO$_3$ and MoO$_3$/MoS$_2$ heterostructures. To minimize perturbation on their 2D morphology, a nonthermal oxidation using O$_2$ plasma was employed. Tohe early stage of the reaction was characterized by the defect-induced Raman peak, drastic quenching of photoluminescence (PL) signals and sub-nm protrusions in atomic force microscopy images. As the reaction proceeded from the uppermost layer to buried layers, PL and optical second harmonic generation signals showed characteristic modulations revealing a layer-by-layer conversion. The plasma-generated 2D oxides, confirmed as MoO$_3$ by X-ray photoelectron spectroscopy, were found to be amorphous but extremely flat with a surface roughness of 0.18 nm, comparable to that of 1L MoS$_2$. The rate of oxidation quantified with Raman spectroscopy decreased very rapidly for buried sulfide layers due to protection by the surface 2D oxides, exhibiting a pseudo-self-limiting behavior. As exemplified in this work, various on-stack chemical transformations can be applied to other 2D materials in forming otherwise unobtainable materials and complex heterostructures, thus expanding the palette of 2D materials building blocks.

Since the first mechanical exfoliation of 2-dimensional (2D) crystalline graphene,[1, 2] various dielectric and semiconducting analogues represented by h-BN and $MoS_2$ have been isolated revealing many of new scientific phenomena and principles.[3-9] Despite being incompatible with mass production of large area samples, this simple method can be applied to essentially any layered materials with weak interlayer interaction,[3] still remaining as one of the best methods for high quality crystals.[10, 11] Since most of single layer (1L) transition metal dichalcogenides ($MX_2$, where M and X denote metal and chalcogen, respectively) were predicted to be thermodynamically stable,[12] the palette of 2-dimensional crystals can be filled with tens of metallic, semiconducting and insulating "colors". In contrast to the "top-down" exfoliation, it was also shown that 2D crystals and their heterostructures can be constructed "bottom-up" from appropriate building blocks via chemical routes as demonstrated by the vapor deposition of graphene and other crystals.[13-17] Alternatively, arbitrarily stacked extended 2D structures can be formed by physically stacking one 2D crystal on top of another exploiting interplanar van der Waals (vdW) interactions.[18] Even with limited "colors" available, there have already been interesting 2D heterostructure-related reports such as tunable metal-insulator transitions,[19] field effect tunneling transistors,[20] photodetectors,[21] *etc*.

Chemical transformation of existing 2D materials may also be useful in further expanding the 2D crystal palette required to realize various functional heterostructures. The top single, few or even all layers of supported 2D materials may be selectively transformed into another of different chemical nature by choosing appropriate chemical reactions. Such "on-stack transformations" may form 2D materials that are not readily generated via the conventional synthetic routes or physical vdW stacking routes. This presents a challenge and opportunity for chemists and materials scientists to explore and develop various chemical reactions that can be applied to each type of 2D crystals. Considering the high aspect ratio (length-to-thickness) of 2D materials, a precise control over the thickness of modified layers further requires a given reaction to be highly uniform across the basal planes. In addition, chemical reactions of 2D materials are also of fundamental significance and deserve systematic investigation since their unique geometric and electronic properties may lead to novel phenomena and findings characteristic of the low dimensional materials. For example, the chemical reactivity of 1L graphene was shown to be much higher than few layer graphene for facile out-of-plane distortion[22-24] or larger susceptibility towards external charge doping.[25] Even the edges of graphene are distinct from the inner area in terms of chemical reactivity[24] and electronic properties.[26] The presence of solid substrates, atomically rough in most cases, has profound effects on the geometric[10, 11, 27] and electronic structures,[28] chemical reactivity[23] and wettability[29] of the supported graphene.

In this regard, the chemical transformation between chalcogenides of transition metals can be a model reaction to explore the chemistry of 2D materials. Whereas the disulfides and diselenides of molybdenum have a strong tendency towards oxidation,[30] the resulting trioxides may also be reduced to dichalcogenides in excess of chalcogenic precursors at elevated temperatures.[31, 32] Atomically thin $MoS_2$ has been well studied drawing a great interest for its direct-indirect bandgap transition[5, 7] depending on the number of layers thus revealing rich photophysics[33] and allowing various optoelectronic applications.[34] $MoO_3$, with a larger bandgap (>2.7 eV), is photochromic[35] and also electrochromic,[36] and thus has been actively investigated for their potentials in displays and smart windows.[37] With a large work function of 6.7 eV, $MoO_3$ films also serve as hole dopants as well as hole transport layers.[38] Due to its high dielectric constant, $MoO_3$ may serve as an ideal gate dielectric[38] for atomically thin transistors based on 2D semiconductors.[39] Although there have been many studies on thin $MoO_3$ films formed by thermal and sputter deposition,[37] their surfaces exhibited a typical roughness larger than a few nm[40] thus far from the layer-by-layer precision that can be routinely achieved with various 2D crystals. Since atomically thin $MoS_2$ layers of high crystallinity or large area can be prepared respectively by the exfoliation[3] or CVD method,[41] 2D oxides of atomic thickness may be generated by chemical transformation of the 2D sulfides.

In this work, we demonstrate $O_2$-plasma-based conversion of 2D $MoS_2$ crystals into highly flat 2D $MoO_3$ or $MoO_3$/$MoS_2$ heterostructures with a layer-by-layer precision. As increasing the plasma oxidation time, the Raman and photoluminescence (PL) intensity from $MoS_2$ decreased indicating loss of the sulfides. X-ray photoelectron spectroscopy (XPS) showed that $Mo^{6+}$ increased at the expense of $Mo^{4+}$, confirming the conversion. Optical second-harmonic generation (SHG) spectroscopy revealed a drastic alternation in intensity, which is explained by breakage and recreation of inversion symmetry in the remaining crystalline sulfides. The plasma-generated 2D oxides were highly flat with a typical roughness of 0.18 nm. The thickness of a single layer oxide was 1.8 ± 0.1 nm, ~2.5 times the interlayer spacing of crystalline α-$MoO_3$. Systematic Raman and atomic force microscopy (AFM) measurements also revealed that the oxidation proceeds from the top layer into the underlying layers with the rate decaying rapidly due to the passivation effect of the surface oxide layers. Our study suggests that similar on-stack chemical transformations can be devised to convert one form of 2D chalcogenide into another or their heterostructures, which should expand the palette of 2D materials.

**Results and Discussion**

**Raman spectroscopic characterization of 2D sulfides and their oxidation:** Figures 1a & 1b show optical micrographs of 1L and 2L $MoS_2$ ($1L_{MS}$ and $2L_{MS}$, respectively) obtained before and after a series of exposures to low-frequency $O_2$ plasma (see Methods for the details). The accumulated exposure time ($t_{ox}$) represents the total reaction time. The optical contrast of both samples significantly decreased at $t_{ox} \geq 26$ s and the flake of $1L_{MS}$ could be barely identified for $t_{ox}$ = 242 s. The change in the optical contrast was attributed to its conversion to 1L $MoO_3$ (1L$_{MO}$) with negligible absorption in the visible range as will be shown below. Similar contrast change was observed for $n$L$_{MS}$ ($n$ = 3 ~ 4) (see Fig. S1). Figure 1c shows the Raman spectra of 1L$_{MS}$ obtained in the ambient conditions for various $t_{ox}$. Among the four Raman-active modes of bulk 2H-MoS$_2$ (space group $D_{6h}^4$),[42] the in-plane $E_{2g}^1$ and out-of-plane $A_{1g}$ were shown to be useful in determining the number ($n$) of layers of thin MoS$_2$ samples.[6] Although their symmetry representations vary depending on $n$ (E' and A$_1$' for odd-numbered $n$L$_{MS}$ belonging to $D_{3h}^1$ space group; E$_g$ and A$_{1g}$ for even-numbered $n$L$_{MS}$ belonging to $D_{3d}^3$ space group),[43] the bulk-notations will be used for simplicity according to the original report for 1L.[6] The frequency difference (Δω) of the two Raman peaks was found to be 18.5 cm$^{-1}$ and their widths were comparable to that of the previous report,[6] which confirms that the sample is single-layered. The lack of the defect-activated LA(M) peak also verifies its high structural quality (Fig. S2).[44] As a more direct proof, atomically resolved transmission electron microscope (TEM) images and diffraction patterns were obtained for freestanding 1L samples (Fig. S3).

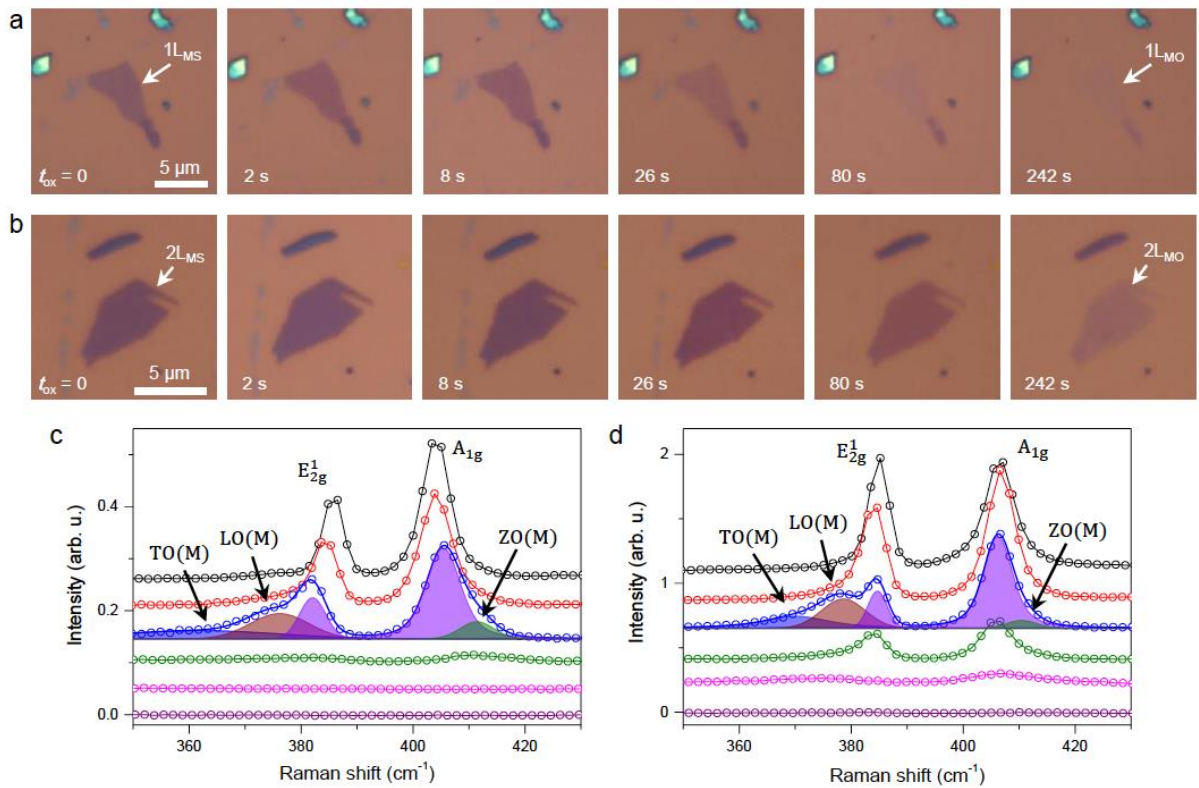

**Figure 1.** Raman spectroscopic characterization of plasma-oxidized MoS$_2$. (a) & (b) Optical micrographs of single and double-layer MoS$_2$ (1L$_{MS}$ & 2L$_{MS}$) obtained after O$_2$ plasma for varying oxidation time ($t_{ox}$). When completely oxidized for 242 s, both samples became transparent indicating formation of corresponding 2D Mo oxides (1L$_{MO}$ & 2L$_{MO}$). (c) & (d) Raman spectra of 1L and 2L MoS$_2$, respectively, obtained with increasing $t_{ox}$ (top to bottom): 0, 2, 8, 26, 80 & 242 s. The raw spectra (open circles and solid lines) were offset for clarity. The blue solid lines for $t_{ox}$ = 8 s are cumulative fits of five Voigt functions representing the color-shaded Raman peaks: two violets for $\mathbf{E_{2g}^1}$ and $\mathbf{A_{1g}}$, one blue peak for TO(M), one wine for LO(M) and one olive for ZO(M) (see the text for their assignment[44]).

Oxidation induces a few distinctive changes in the Raman spectra. The in-plane Raman mode became softened by ~1 cm$^{-1}$ on the first exposure for 2 s and downshifted further with the other mode moving in the opposite direction when subjected to additional exposure ($t_{ox}$ = 8 s). Since the newly evolving Raman peaks have respectively lower and higher peak frequencies than their pristine counterparts, their frequency difference, $\Delta\omega_{(NC)}$, is ~5 cm$^{-1}$ larger than that of pristine 1L$_{MS}$. Both peaks also became broadened asymmetrically and attenuated in intensities. For $t_{ox}$ = 26 s which induced significant loss in the optical contrast in Fig. 1a, for example, the Raman intensities, $I(E_{2g}^1)$ and $I(A_{1g})$ decreased to ~5% of those for the pristine sample with both peaks moving further away from each other and further broadened. The additional exposure ($t_{ox}$ = 80 s) lead to no detectible signal for both peaks, indicating complete loss of MoS$_2$. Within the phonon confinement model, the oxidation-induced downshift (upshift) of $E_{2g}^1$ ($A_{1g}$) can be explained by the relaxation of the fundamental Raman selection rule.[45] In its essence, the defects confining phonons allow scattering of phonons near the zone center with energies lower (higher) than $E_{2g}^1$ ($A_{1g}$). The low (high)-frequency shoulders of $E_{2g}^1$ ($A_{1g}$) that caused the asymmetric broadening are also due to the defects, which selectively allow scattering of the phonons near the M points in the double-resonance Raman scattering mechanism.[46] Indeed the spectra of partially oxidized 1L and 2L (Fig. 1c & 1d) are well described by the five Voigt functions including the defect-activated TO(M), LO(M) and ZO(M).[44, 45] As discussed in detail below, plasma-generated reactive oxygen species create sub-oxide defects (MoS$_x$O$_{1-x}$) on the basal plane of MoS$_2$, essentially forming a nano-crystalline (NC) phase MoS$_2$ sheet.

The Raman spectra of multi-layered samples showed similar spectral changes also with a layer-by-layer oxidation behavior. The in-plane mode of 2L$_{MS}$ in Fig. 1d, for example, showed a noticeable broadening due to the M-point phonons for $t_{ox}$ = 8 s and became sharper with 40% attenuated intensity for $t_{ox}$ = 26 s. An additional exposure ($t_{ox}$ = 80 s) lead to further decrease in intensity and significant broadening. The NC-phase observed for $t_{ox}$ = 8 s is formed on the top MoS$_2$ layer with the bottom layer remained intact. The disappearance of the NC-phase features for $t_{ox}$ = 26 s can be attributed to complete oxidation of the top layer, which is only weakly coupled with the still pristine bottom one responsible for the remaining half of the Raman signals. The second appearance of the broadening for $t_{ox}$ = 80 s indicates that the oxidative attack reached the bottom layer forming NC-phase domains. An additional exposure ($t_{ox}$ = 242 s) lead to a complete loss of Raman signal. These observations indicate that the oxidation proceeds in a layer-by-layer manner and is much slow for inner layers that are protected by the outer oxides.

**Symmetry characterization by SHG:** Since the optical SHG requires the lack of an inversion symmetry,[47] it can serve as a sensitive symmetry probe for chemical modification occurring in thin MoS$_2$. As shown in Fig. 2a, a strong SHG signal occurred at 400 nm when the fundamental 800-nm Ti:sapphire laser beam with a nominal pulse width of 140 fs was irradiated on non-centrosymmetric pristine 1L$_{MS}$.[48] An input power dependence of the SHG peak intensity ($I_{SHG}$) confirmed the two-photon process (Fig. S4). The intensity of the SHG peak ($I_{SHG}$) indeed decreased by the plasma oxidation and finally reached the zero level for 1L$_{MO}$ ($t_{ox}$ = 20 s), which correlates nicely with the Raman intensity variation of MoS$_2$ (Fig. 2b). Thus the attenuation of $I_{SHG}$ can be attributed to destruction of 1L MoS$_2$ crystal. As shown in Fig. 2a, 2L MoS$_2$ presents a striking contrast to the case of 1L. $I_{SHG}$(2L$_{MS}$) is ~60 times smaller than $I_{SHG}$(1L$_{MS}$) since even-number-layered MoS$_2$ crystals are centrosymmetric and thus do not support SHG. The residual signal of 2L$_{MS}$ is attributed to a minor interlayer asymmetry induced by the presence of the SiO$_2$/Si substrate.[49] A series of oxidation treatments, however, activated and then deactivated SHG by the 2L flake (Fig. 2a). Figure 2b reveals that $I_{SHG}$ intensity increased at the expense of the Raman signal represented by $I(A_{1g})$. Since $I_{SHG}$ reached its maximum for $t_{ox}$ = 30 s for which the Raman intensity is close to that of 1L$_{MS}$, the treated sample essentially corresponds to 1L$_{MO}$/1L$_{MS}$. This assignment is also supported by the fact that additional treatments lead to a decrease in $I_{SHG}$ indicating oxidative degradation of the bottom sulfide layer. The lack of SHG signal from the completely oxidized layers (1L$_{MO}$ and 2L$_{MO}$) might be explained if the 2D oxides are a centrosymmetric crystal like bulk α-MoO$_3$ belong to a space group of $D_{2h}^{16}$.[50] As will be shown below, however, the 2D oxides are far from α-MoO$_3$ and more likely to be amorphous and isotropic not generating SHG signal.

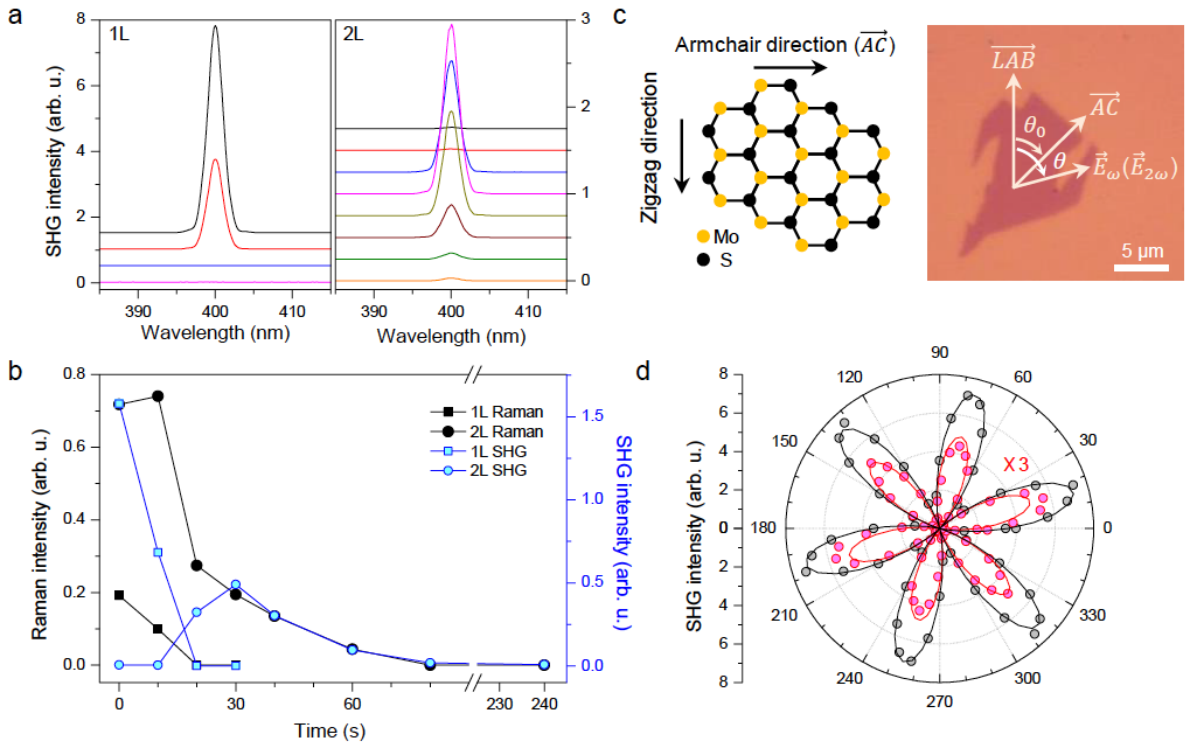

**Figure 2**. Nonlinear optical characterization of plasma-oxidized MoS$_2$. (a) Optical SHG spectra of 1L and 2L MoS$_2$ obtained as increasing $t_{ox}$ (top to bottom): 0, 10, 20, 30, 40, 60, 80 & 240 s. (b) Raman and SHG intensities of 1L and 2L MoS$_2$ given as a function of $t_{ox}$. The maximum intensity of 2L for $t_{ox}$ = 30 s amounts to ~1/3 of that for the pristine 1L. (c) Schematic view (left) of MoS$_2$ lattice with armchair and zigzag edges and an optical image (right) of 1L MoS$_2$ used in the SHG measurements: With the polarization directions of the input ($\vec{E}_\omega$) and output ($\vec{E}_{2\omega}$) fixed, the sample with a predefined laboratory axis ($\overrightarrow{LAB}$) was rotated by $\theta$. The armchair direction ($\overrightarrow{AC}$) forms an angle ($\theta_0$) with respect to $\overrightarrow{LAB}$. (d) Polar plot of the parallel SHG intensity ($I_{SHG}^{\parallel}$) of pristine (gray) and half-oxidized (magenta) 1L MoS$_2$ given as a function of $\theta$. Using the fact that $I_{SHG}^{\parallel} \propto cos^2 3(\theta - \theta_0)$ which is represented by the solid lines, the armchair direction of the sample is determined with $\theta_0$ = 15.0 ± 0.2 degree.

Despite the degradation of the crystalline lattice, the crystallographic orientation could still be determined for partially oxidized MoS$_2$ since $I_{SHG}$ is strongly dependent on the polarization angle ($\theta - \theta_0$) between the fundamental polarization ($\vec{E}_\omega$) and armchair direction ($\overrightarrow{AC}$) defined in Fig. 2c. For polarization-resolved SHG measurements, samples were rotated to vary $\theta$ and the SHG signal parallel to the input polarization, $I_{SHG}^{\parallel}$, was collected using a polarizer located in front of the detector. The polar plot for 1L$_{MS}$ in Fig. 2d revealed the 6-fold symmetry of $I_{SHG}^{\parallel} \propto \cos^2 3(\theta - \theta_0)$,[48] which also determined that $\theta_0$ = 15.0 ± 0.2 degree, the angle of $\overrightarrow{AC}$ with respect to a preset laboratory axis ($\overrightarrow{LAB}$). When oxidized to give ~0.4 L$_{MS}$ judged from its Raman spectra ($t_{ox}$ = 15 s), $I_{SHG}^{\parallel}$ decreased by ~75% but still showed the same angular dependence as the pristine 1L$_{MS}$. The oxidation-induced modulation in $I_{SHG}$ (Figs. 2a and 2b) and robust $\theta$-dependence demonstrates the SHG process can be utilized in characterizing chemical changes in 2-dim materials.

**Formation of ultraflat 2D oxides with a single-layer precision:** The on-stack transformation was found to proceed in a layer-by-layer manner and could be controlled with a high thickness resolution even allowing formation of single-layer MoO$_3$. To monitor morphological changes by the transformation, an amplitude-modulated non-contact AFM was exploited. The height profile across the pristine 2L$_{MS}$-1L$_{MS}$ area in Fig. 3a revealed a step with a height of 0.70 ± 0.1 nm, which agrees well with the interlayer spacing (0.62 nm) of 2H-MoS$_2$.[51] Note that the height of the 1L$_{MS}$-SiO$_2$ step varied in the range of 1.2 ~ 1.8 nm from sample to sample due to the chemical and electrostatic contrast between dissimilar materials[52, 53] and interlayer molecular species trapped during the exfoliation. When 2L$_{MS}$ was completely oxidized ($t_{ox}$ = 242 s), the 2L$_{MO}$-1L$_{MO}$ step height increased to 1.8 ± 0.1 nm (Fig. 3b & 3e). As depicted in Fig. 3f, the step height corresponds to the thickness of a single-layer MoO$_3$ and is ~2.5 times the interlayer spacing (0.69 nm) of α-MoO$_3$.[50] A similar change was observed for the 2L$_{MS}$-1L$_{MS}$ step of another sample when treated for $t_{ox}$ = 242 s (Fig. 3c & 3d). After the same treatment, however, the 3L$_{MS}$-2L$_{MS}$ step height changed only by ~0.2 nm. In contrast to the 2L region, the Raman spectrum of the 3L region showed that ~15% of the lowermost layer still remained intact (see Fig. S5 for Raman spectra obtained after each step of the sequential oxidations). Thus, the step height measured across the 3L-2L region corresponds to the thickness of the partially oxidized bottom layer.

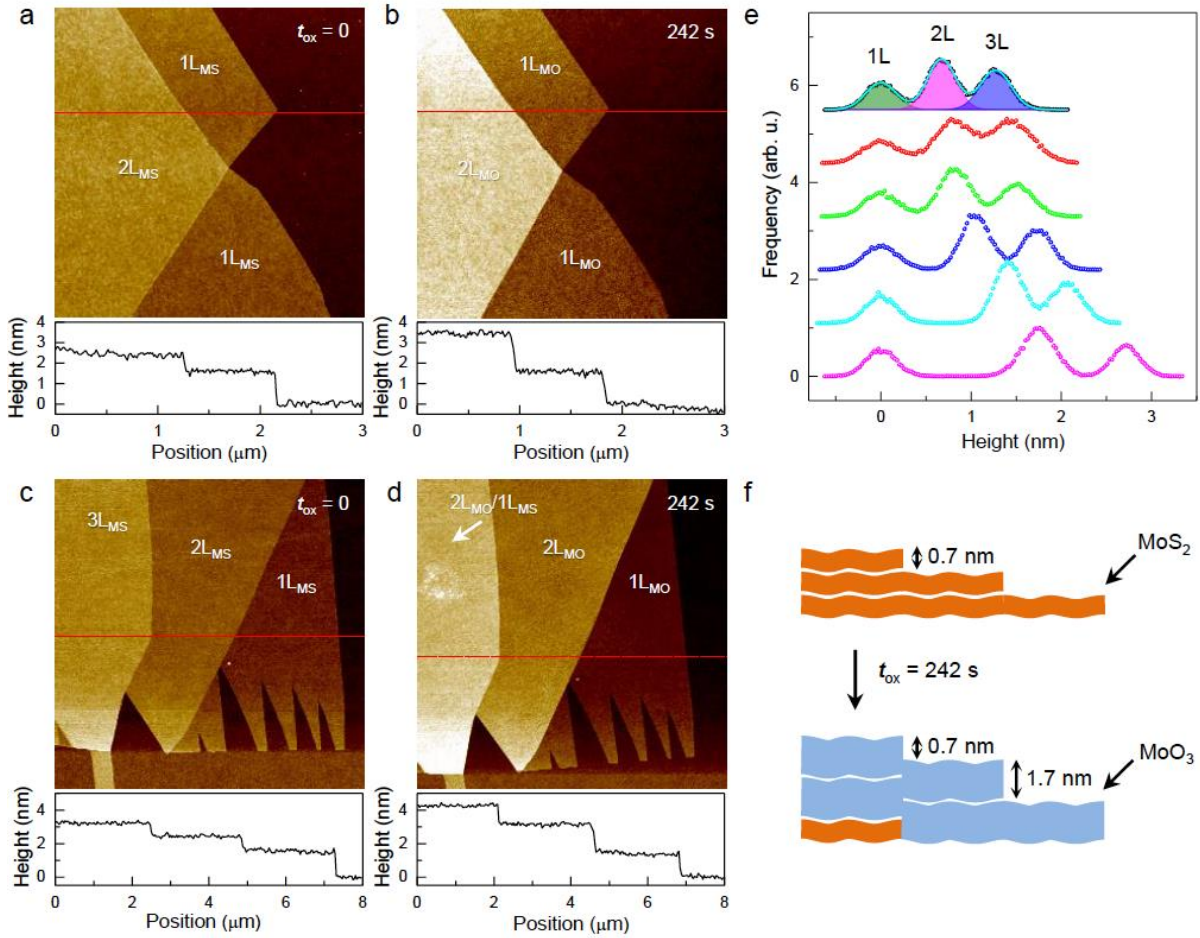

**Figure 3.** Layer-by-layer oxidation of few-layer MoS$_2$. (a) & (b) Non-contact AFM height images of a 1L$_{MS}$-2L$_{MS}$ sample obtained before and after oxidation ($t_{ox}$ = 242 s), respectively. Height profiles across the red lines in (a) and (b) are also shown below the AFM images. (c) & (d) Non-contact AFM height images of another sample with 1L$_{MS}$, 2L$_{MS}$ and 3L$_{MS}$ regions before and after oxidation ($t_{ox}$ = 242 s), respectively. Height profiles across the red lines in (c) and (d) are also shown below the AFM images. (e) Height histograms of the sample in (c & d) obtained for the 1L, 2L and 3L areas as a function of $t_{ox}$ (top to bottom): 0, 2, 8, 26, 80 & 242 s. The mean height of 1L area was set to be zero. Root-mean-square roughness (R$_q$) of each area was determined by fitting the data with 3 color-shaded Gaussian functions as shown for the pristine sample (see Fig. S6 for the variation of the roughness induced by the oxidation). (f) Schematic diagram of "top-down" layer-by-layer oxidation that converts MoS$_2$ into MoO$_3$.

Despite the increase in the thickness, MoO$_3$ layers were still highly flat as can be seen in the AFM height images in Fig. 3. The change in the surface morphology of the 2D oxides was further quantified by the root-mean-square roughness (R$_q$) or standard deviation of height distribution shown in Fig. 3e. R$_q$ of 1L$_{MS}$, 170 ± 10 pm which is slightly higher than that of graphene,[54] is equivalent to that of completely oxidized 1L$_{MO}$, 180 ± 10 pm for $t_{ox}$ ≥ 80 s as shown in Fig. S6. Notably, however, the surface roughness showed a spike of 230 ± 10 pm for $t_{ox}$ = 2 s and a sharp decrease for $t_{ox}$ = 8 s. Similar spikes were also observed for $2L_{MS}$ and $3L_{MS}$ when treated for 2 s. $R_q$'s of $2L_{MO}$ and $3L_{MO}$ (170 ± 10 pm) are also equivalent to those of their pristine counterparts. We attribute the roughness surge to clusters of partially oxidized $MoS_2$, possibly molybdenum oxysulphides ($MoS_xO_y$)[55] which serve as nucleation sites during the plasma oxidation. This picture is also corroborated by the drastic PL quenching for $t_{ox}$ = 2 s as will be shown below. However, the Raman spectra did not show a noticeable change (Fig. 1c & 1d) since majority of the sample remained intact for the short reaction time.

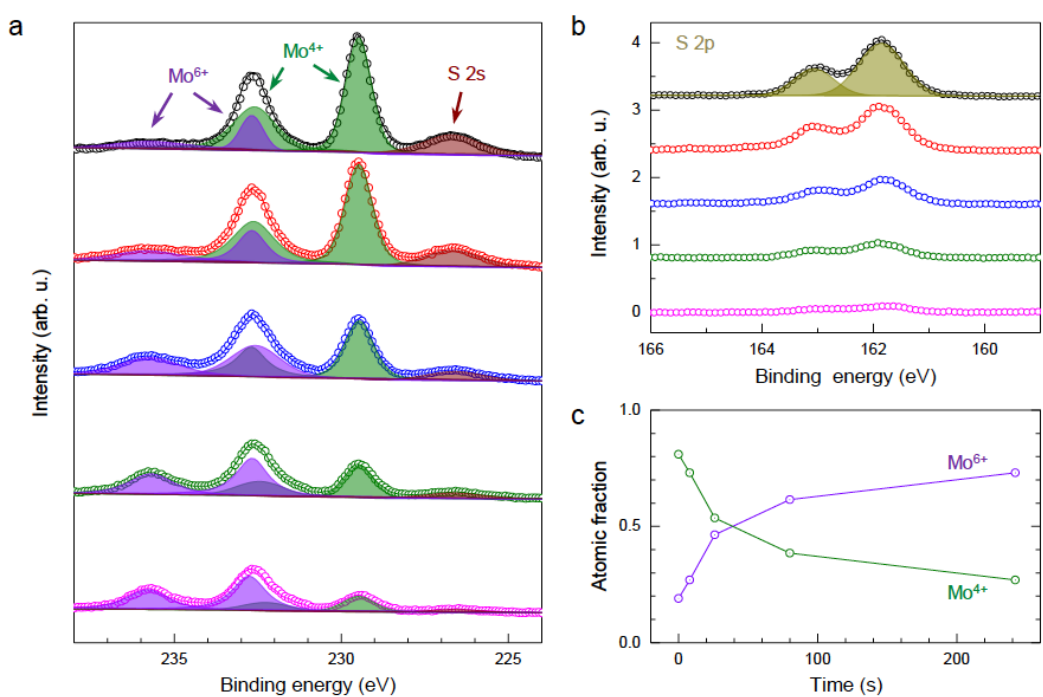

**Figure 4.** Elemental analysis of plasma-oxidized $MoS_2$. (a) Mo 3d XPS spectra of CVD-grown $2L_{MS}$ samples oxidized for various $t_{ox}$ (top to bottom): 0, 2, 8, 26, 80 & 242 s. The spectra were fitted with multiple mixed Lorentzian-Gaussian functions. The green shades correspond to $Mo^{4+}$ ($3d_{5/2}$) and $Mo^{4+}$ ($3d_{3/2}$) peaks, and the violet shades to $Mo^{6+}$ ($3d_{5/2}$) and $Mo^{6+}$ ($3d_{3/2}$) peaks. At the binding energy of ~227 eV, S 2s peak can be observed. (b) S 2p XPS spectra of the same samples shown in (a). The spectra were also fitted with two sub-peaks, $2p_{1/2}$ and $2p_{3/2}$. (c) The overall photoelectron intensity of Mo species and atomic fractions of $Mo^{4+}$ and $Mo^{6+}$ given as a function of $t_{ox}$.

**Chemical nature of 2D oxides:** The chemical transformation was revealed by the X-ray photoelectron spectroscopy (XPS). Because the probe size of the exploited XPS instrument was larger than the typical size (< 20 µm) of exfoliated samples, large-area $MoS_2$ films grown by the chemical vapor deposition (CVD) were exploited (see Methods for the details of preparation). Whereas the effective thickness was found to be 2L based on the peak difference (Δω), their Raman intensity was lower than that of mechanically exfoliated samples, suggesting lower crystallinity (Fig. S7). Figure 4a presents the Mo 3d spectra obtained as a function of $t_{ox}$. The pristine sample showed the strong doublet of Mo$^{4+}$ arising from MoS$_2$ with a binding energy ($E_B$) of 229.5 eV (232.6 eV) for 3d$_{5/2}$ (3d$_{3/2}$), which is consistent with the average literature value of $E_B$(3d$_{5/2}$) = 229.25 eV for bulk MoS$_2$ crystals.[56] The Mo$^{6+}$ species of the pristine sample, responsible for the minor doublet with $E_B$ = 232.7 eV (235.7 eV) for 3d$_{5/2}$ (3d$_{3/2}$), were attributed to residual MoO$_3$ (< 20% of the total Mo species) formed before or after the CVD growth. Upon the plasma treatment, the Mo$^{6+}$ doublet significantly grew at the expense of the Mo$^{4+}$ doublet indicating the conversion to MoO$_3$. The oxidation was also confirmed by the decrease in the intensities of the S 2s peak ($E_B$ = 226.7 eV) in Fig. 4a and S 2p doublet ($E_B$ = 161.9 eV for 2p$_{3/2}$) in Fig. 4b. The current data cannot exclude the possibility that a small fraction of Mo$^{5+}$ species ($E_B$ = 231.1 eV for 3d$_{5/2}$)[57] representing oxygen vacancy defects generated by incomplete oxidation may be present.

To shed more light on the chemical transformation, we obtained optical absorption spectra using the reflectance spectroscopy in Fig. 5 (see Methods for the details). For a very thin film (thickness $\ll \lambda$) supported on a thick transparent substrate with refractive index of n$_0$, the fractional change in reflectance ($\delta_R$) is given as follows: $\delta_R = \frac{R-R_0}{R_0} = \frac{4}{n_0^2-1}A$, where R, R$_0$ and A are reflectance of the film, reflectance of the bare substrate, and absorbance.[58] Pristine 1L$_{MS}$ ~ 3L$_{MS}$ exfoliated on quartz substrates showed characteristic absorption peaks at 1.90, 2.05 and 2.85 eV, which were denoted respectively by A, B and C according to an early work.[59] The two former excitonic peaks originate from the direct-gap transitions between the valence and conduction bands at the K points in the Brillouin zone.[5] The high energy peak, C, arises from nearly degenerate multiple excitonic states[60] or transitions across nested valence and conduction bands.[61] All the three peaks downshifted in energy with increasing thickness.[5] When the samples were treated for $t_{ox}$ = 80 s which is expected to oxidize ~1.7L, the overall absorbance decreased significantly for all the thicknesses. The A & B peaks of 1L, in particular, almost disappeared with the major absorption edges relocated to > 2.7 eV, which agrees well with the fact that MoO$_3$ is a wide-bandgap semiconductor.[62] The partially oxidized 2L exhibited downshifts of A and B peaks but an upshift of C peak, the degree of which were much less for 3L. The spectral changes are attributed to the oxidation-induced defects and can be explained by the drastic changes in the electronic band structures of O-substituted $MoS_2$.[63] To determine the optical bandgap of the 2D $MoO_3$, the Tauc plot analysis was taken for the UV/Vis absorption spectra of CVD-grown large-area samples (Fig. S8). Despite some inconsistencies in the literature,[64] it is generally accepted that an optical gap energy ($E_g$) of an amorphous semiconductor can be extrapolated using $(\alpha h\nu)^{1/m} \propto (h\nu - E_g)$,[65, 66] where $\alpha$, $h\nu$ and m are respectively absorption coefficient, photon energy and exponent specifying the nature of the optical transition (m = 2 for an indirect allowed transition; m = 1/2 for a direct allowed transition). By choosing m = 2 for α-$MoO_3$ is an indirect bandgap semiconductor,[67] $E_g$ was determined to be ~3.0 eV for $1L_{MO}$ and $2L_{MO}$. The resulting bandgap energies are in a good agreement with those of α-$MoO_3$ crystals[62] and thermally deposited $MoO_3$ films.[68, 69] Despite the similar bandgap energies and apparent layered structure of plasma-generated $MoO_3$, however, no Raman signal could be detected from the 2D oxides possibly due to their amorphous nature. A careful Raman analysis in comparison with mechanically exfoliated thin α-$MoO_3$ crystals set an upper bound for α-$MoO_3$ content ~1% (Fig. S9).

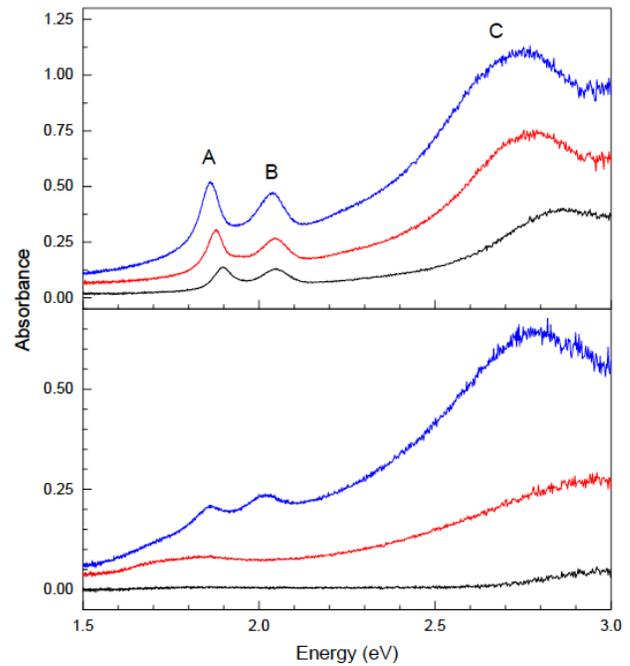

**Figure 5.** The absorption spectra of 1L ~ 3L MoS$_2$ exfoliated onto quartz substrates: (top) before and (bottom) after oxidation for $t_{ox}$ = 80 s. The absorbance was determined from optical reflectance contrast measurements (see Methods for the details). See the text for the origins of the three prominent peaks denoted by A, B and C. For comparison with UV/Vis transmission measurements for CVD-grown samples and estimation of their optical bandgaps using Tauc plots, see Fig. S7.

**PL intensity modulation:** The change in the photoluminescence (PL) spectra in Fig. 6 revealed further details of the chemical modifications occurring in $MoS_2$ thin membranes. The PL spectrum of the pristine $1L_{MS}$ mainly consists of the two exciton peaks, A and B, respectively located at 1.84 and 2.00 eV, which are in a good agreement with the absorption spectra. Both originate from the direct-gap transitions between the conduction and valence bands at the K points in the Brillouin zone.[5, 7] First, we note that the PL is drastically quenched by the plasma treatments. After 2-s exposure, for example, the intensity of A ($I_A$) decreased by ~65% although the Raman spectra in Fig. 1 showed no significant change in $I(E_{2g}^1)$ and $I(A_{1g})$. The quenching can be attributed to the plasma-generated defects that were confirmed by the defect-induced Raman peaks, as will be further discussed below. When further defects were introduced to form the NC phase ($t_{ox}$ = 8 s), the PL intensity was only ~10% of the pristine value. The completely oxidized $1L_{MO}$ gave no PL signal, which is consistent with the fact that its $E_g$ is larger than the excitation photon energy of 2.41 eV. $2L_{MS}$ in Fig. 6b showed a similar defect-mediated PL quenching: ~60% decrease for $t_{ox}$ = 2 s and almost zero PL signal for $t_{ox}$ = 8 s which introduced significant amount of defects judged from the Raman spectrum (Fig. 1c). Upon further oxidation ($t_{ox}$ = 26 s), however, the PL intensity increased back to ~30% of that for the pristine $2L_{MS}$. We note that the Raman spectra showed that the top layer was almost completely oxidized with the bottom layer mostly intact (Fig. 1d). This suggests that the system essentially became a $1L_{MO}/1L_{MS}/SiO_2$ sandwich, which further corroborates the layer-by-layer oxidation. Thicker samples showed a similar trend but the PL intensity oscillation of $nL_{MS}$ could be seen only when $t_{ox}$ was selected in such a way that $(n-1)L_{MO}/1L_{MS}$ was nicely formed.

Structural defects in crystalline semiconductors have various modes of interactions with excited charge carriers, and their interactions become more significant in lower dimensional materials due to confined wave functions within a tighter space. For carbon nanotubes with low lying dark exciton states, $sp^3$-type defects form new sets of radiative energy levels endowing an enhanced PL quantum yield.[70] Defects may trap excited free charge carriers and mediate ultrafast Auger decay of photoexcitation in quantum dots.[71] Excitons can be localized at defects and radiatively decay at lower energies as shown for 1L $MoS_2$ irradiated with α particles[72] and 1L $WS_2$ treated with Ar plasma.[73] Defects in some 2D semiconductors were also shown to serve as single photon emitters.[74, 75] Localization followed by non-radiative decay is one among various quenching routes of excitons, which also include exciton-exciton annihilation and electron-phonon interactions.[76] The PL quenching observed at the very early stage of the oxidative conversion can be attributed to partially oxidized defects including the nanometer scale protrusions that were responsible for the roughness spikes (Fig. S6). The higher defect-sensitivity of PL can be more clearly seen in the comparison between Raman and PL intensities given as a function of $t_{ox}$ (Fig. S10).

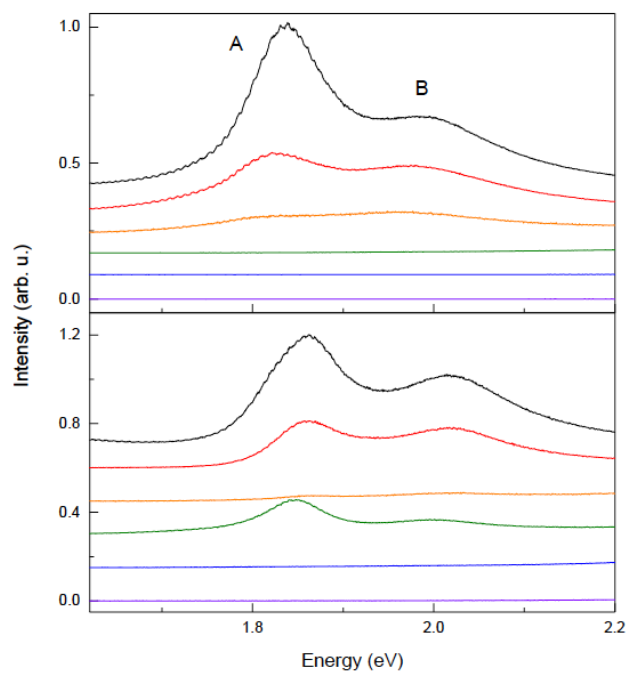

**Figure 6.** Photoluminescence spectra of 1L (top) and 2L (bottom) $MoS_2$ for varying $t_{ox}$ (top to bottom): 0, 2, 8, 26, 80 & 242 s. The two emission peaks denoted by A and B have the same origins as the corresponding absorption peaks in Fig. 5.

**Layer-by-layer conversion:** The overall change can be summarized by the scheme for $3L_{MS}$ with terraces of $2L_{MS}$ and $1L_{MS}$ presented in Fig. 3f. At the early stage of reactions ($t_{ox}$ = 2 s), the top surface of $3L_{MS}$ exposed to the gaseous oxidants undergoes partial oxidation forming nanometer-scale oxysulfides ($MoS_xO_y$) clusters that are responsible for the increased $R_q$. At this stage, the average thickness of the top layer is not much different from that of pristine $1L_{MS}$ or the interlayer spacing of 2H-$MoS_2$ crystals. Although the Raman spectra reveal no significant change, the PL intensity of $3L_{MS}$ drops significantly due to the defects serving as PL quenchers. Additional exposure to oxidants induces the NC-phase Raman peaks and almost complete quenching of the PL. When the first top layer is completely converted to $MoO_3$, its thickness increases to 1.8 nm which is 2.5 times the interlayer spacing of α-$MoO_3$ crystals. At this stage where $1L_{MO}/2L_{MS}$ is formed, the PL intensity can be recovered back to that of $2L_{MS}$ because of lack of quenching sites within the remaining $2L_{MS}$. Further exposure will repeat the above reactions for the second top layer but on a much decreased reaction rate.

Although the Raman analysis suggests that the 2D oxides are amorphous, their high flatness and modulation in SHG and PL intensities indicate that the oxidation reaction proceeds in the layer-by-layer manner. Thus, it is likely that the oxides are also layered like α-MoO$_3$. Since the number density of Mo in a single α-MoO$_3$ layer[67] is 18% higher than that for 1L$_{MS}$,[51] a 2D conversion into the crystalline oxide would lead to a large tensile strain (~9%) when the areal density of Mo is maintained or generate voids requiring significant mass transport to compensate the difference in the Mo densities. Moreover, the average thickness of 1L$_{MO}$ (Fig. 3) is 2.5 times the interlayer spacing of α-MoO$_3$. These facts suggest that the 2D oxides are much less dense than α-MoO$_3$. Despite the apparent flatness and lack of pits or cracks in the 2D oxides, these considerations lead us to a conclusion that there must be numerous structural irregularities including fine voids that cannot be detected by the AFM probes. Our results also showed that the plasma oxidation is highly effective for the topmost sulfide layer and increasingly slower for the next buried layers. Such voids may serve as a passage route for the oxidants required for the reactions underneath the top layer.

In summary, we demonstrated a model 2D chemical reaction that converts 1L sulfides into 1L oxides thus allowing MoO$_3$/MoS$_2$ heterostructure. Single and few-layer MoS$_2$ prepared via micromechanical exfoliation of 2H-MoS$_2$ crystals were treated with strong oxidants generated from O$_2$ plasma. The early stage of the reaction was detected by the defect-induced Raman peaks, drastic quenching of PL signals and sub-nm protrusions in AFM images. As the reaction proceeded from the uppermost layer to buried layers, PL and SHG signals showed characteristic modulations revealing a layer-by-layer conversion. The plasma-generated 2D oxides, confirmed as MoO$_3$ by XPS, were found to be amorphous but highly flat with a surface roughness of 0.18 nm, comparable to that of 1L MoS$_2$. The rate of oxidation quantified with Raman spectroscopy decreased very rapidly for buried sulfide layers due to protection by the surface 2D oxides. As exemplified in this work, the on-stack chemical transformation can be applied to other 2D materials in forming otherwise unobtainable materials and complex heterostructures, thus expanding the palette of 2D materials building blocks.

**Experimental Section**

**Mechanical exfoliation.** Single and few-layer MoS$_2$ samples were prepared by mechanical exfoliation[3] of molybdenite, a natural mineral of 2H-MoS$_2$ (SPI). Silicon wafers with 285-nm-thick SiO$_2$ layers were used as substrates. Ultrathin layers of MoS$_2$ were identified using an optical microscope and their numbers of layers and quality were determined using Raman spectroscopy.[6] To prepare thin α-MoO$_3$ crystalline flakes as a Raman standard (Fig. S9), a similar mechanical exfoliation was applied to a powder of MoO$_3$ (Materion).

**Growth of large-area samples by CVD.** Two different CVD growth methods were used in synthesizing large area MoS$_2$ films. In one approach for the samples used for XPS measurements, Mo metal of 0.5 nm for 2L$_{MS}$ was deposited on SiO$_2$/Si substrates by e-beam evaporation. The samples were positioned in a quartz tube furnace, which was then evacuated to a low vacuum. The samples were heated to 750 °C under flow of Ar at a rate of 50 mL/min. Since the surface of deposited Mo films usually becomes oxidized when exposed to the ambient air during transfer to the furnace, a H$_2$ gas was briefly introduced at 750 °C to reduce the Mo oxides. After the pre-annealing and reduction processes, a H$_2$S/H$_2$/Ar gas mixture (1:5:50) was introduced for 15 min to sulfurize Mo films into MoS$_2$. The pressure in the furnace was maintained at 300 mTorr during the sulfurization step. To enhance the crystallinity of grown films, the samples were further annealed briefly at 1000 °C under a flow of H$_2$S/Ar gas mixture (1:50). In the other approach for the samples used UV/Vis measurements, 5 ~ 10 mg of MoO$_3$ powder (Sigma-Aldrich) in a quartz boat was placed at the center of the furnace and quartz substrates were placed near the boat in the downstream. The furnace was heated up to 600 °C at a rate of 20 °C/min under an Ar gas with a flow rate of 200 mL/min. At 600 °C, a H$_2$S gas was introduced at a rate of 1 mL/min to sulfurize MoO$_3$ into MoS$_2$ for 30 min. Then the samples were rapidly cooled down to room temperature. The average thickness and quality of grown films were characterized by their Raman and PL spectra.

**Plasma oxidation.** For oxidation, samples were treated with either of the two low-frequency plasma instruments operated at 50 kHz, a commercial unit (Femto Science Inc., Cute-1MP) and a quartz-tube-type unit. The partial pressure of O$_2$ in the two plasma chambers was 540 and 300 mTorr, respectively. For oxidation reactions, oxygen plasma was maintained for a pre-specified period of $t_{ox}$ at a power of 10 W. Because of the differences in the instruments and detailed procedures, the apparent oxidation rate was ~3 times larger for the former than the latter. Thus, $t_{ox}$ given in this work has been corrected for the difference.

**Raman and PL Measurements.** The Raman and PL spectra of the samples were obtained with a microscope-based (Nikon, Ti) Raman setup that is equivalent to what detailed elsewhere.[77, 78] Briefly, the 514 nm excitation beam from a solid state laser (Cobolt, Fangdango) was focused onto a spot of ~1 μm in diameter using an objective lens (40X, numerical aperture = 0.60). The back-scattered signal was collected by the same objective and guided to a spectrograph (Princeton Instruments, SP2300) combined with a liquid nitrogen-cooled charge-coupled detector (CCD) (Princeton Instruments, PyLon). The spectral resolution judged from the FWHM of the Rayleigh line was 3.0 and 12 cm$^{-1}$ for Raman and PL spectra, respectively.

**SHG measurements.** A similar microscope-based spectroscopy setup was employed for the SHG detection. A train of 140-fs pulses from a Ti:Sapphire laser (Coherent, Chameleon) operated at 800 nm was focused onto a spot of 1.6 μm in FWHM with an objective lens (40X, numerical aperture = 0.60). The backscattered SHG signal centered at 400 nm was collected and fed to a spectrograph (Andor, Shamrock 303i) equipped with a thermoelectrically cooled CCD (Andor, Newton). To vary the polarization angle of the fundamental laser with respect to the MoS$_2$ lattice vectors, samples were rotated in a rotational mount with an angular accuracy better than 0.2 degree. For a polarized detection, the SHG signal was filtered with a polarizer located in front of the spectrograph.

**AFM measurements.** The topographic details of the samples were investigated by using an atomic force microscope (Park Systems, XE-70). The height images were obtained in a non-contact mode using Si tips with a nominal tip radius of 8 nm (MicroMasch, NSC-15). The AFM tip was driven to a free oscillation with an amplitude of 20 nm and engaged in amplitude-modulated scanning with an amplitude set-point of ~14 nm.

**XPS measurements.** The elemental information of the CVD-grown samples was obtained by using an X-ray photoelectron spectrometer (Thermo Scientific, K-Alpha$^{TM+}$). XPS measurements were carried out with Al K$_\alpha$ line (1.4866 keV). The binding energy of the photoelectrons was calibrated with respect to Mo(IV) 3d$_{5/2}$ ($E_B$ = 229.5 eV) which was more reliable than C 1s peak ($E_B$ = 284.6 eV) originating from carbon residues on the samples.

**Optical absorption measurements.** For the samples exfoliated onto quartz substrates, the fractional changes in reflectance were used to obtain absorption spectra. As a broadband Vis/NIR light source, the output of a tungsten-halogen lamp (iiSM Inc., Mighty Light Beam) was collected with a multimode optical fiber (core diameter of 50 μm) and guided to the micro-spectroscopy setup exploited for the SHG measurements. The FWHM of the focus spot was 0.8 and 2.0 μm in the visible and NIR ranges, respectively. For the samples CVD-grown on quartz substrates, the absorption spectra were obtained with a UV/visible spectrometer (Jasco, V-650).

**TEM measurements.** For TEM measurements, MoS$_2$ flakes exfoliated onto SiO$_2$/Si substrates were transferred to carbon-film grids with 2-μm holes (Quantifoil) using isopropyl alcohol and KOH solution.[79] The samples were analyzed using an aberration-corrected FEI Titan Cubed TEM (FEI, Titan$^3$ G2 60-300), which was operated at 80 kV acceleration voltage with a monochromator. The microscope provides sub-angstrom resolution at 80 kV and -11 ± 0.5 μm of spherical aberration (C$_s$).

ASSOCIATED CONTENT

**Supporting Information.** A. Optical contrast of plasma-oxidized MoS$_2$; B. Defect-activated scattering of LA(M) in plasma-oxidized MoS$_2$; C. TEM characterization of freestanding 1L MoS$_2$; D. Power dependence of SHG signal from 1L MoS$_2$; E. Thickness-dependent Raman characterization of plasma-oxidized MoS$_2$; F. Effect of O$_2$ plasma treatment on surface roughness of MoS$_2$; G. Comparison of CVD-grown and exfoliated MoS$_2$ using Raman spectroscopy; H. Optical bandgap of plasma-generated MoO$_3$; I. Possible content of α-MoO$_3$ in plasma-generated oxides; J. Oxidation-induced decays in Raman and PL signals of 1L ~ 4L MoS$_2$.

AUTHOR INFORMATION

**Corresponding Author**

*E-mail: sunryu@postech.ac.kr

ACKNOWLEDGMENT

This work was supported by the Center for Advanced Soft-Electronics funded by the Ministry of Science, ICT and Future Planning as Global Frontier Project (NRF-2014M3A6A5060934) and also by the National Research Foundation of Korea (NRF-2015R1A2A1A15052078).